\documentclass[twocolumn,showpacs]{revtex4}
\usepackage{graphicx}
\begin{document}
\draft
\title{Spin gap behavior in Cu$_2$Sc$_2$Ge$_4$O$_{13}$ by $^{45}$Sc nuclear
magnetic resonance}
\author{ C. S. Lue$^{1,}$\footnote{Email:
cslue@mail.ncku.edu.tw}, C. N. Kuo$^{1}$, T. H. Su$^1$, and G. J.
Redhammer$^{2,}$\footnote{Email: guenther.redhammer@aon.at}
 }
%\author{C. S. Lue,$^{1,}$\cite{csl} C. N. Kuo,$^1$ T. H. Su,$^1$ and G. J. Redhammer$%
%^{2,}$\cite{gjr}}
\affiliation{ $^{1}$Department of Physics, National Cheng-Kung
University, Tainan 701, Taiwan \\
$^2$Department of Material Science, Division of Mineralogy, University of\\
Salzburg, Hellbrunnerstr. 34, Salzburg A-5020, Austria
 }
%\address{$^1$Department of Physics, National Cheng Kung University, Tainan 70101,\\
%Taiwan}
%\address{$^2$Department of Material Science, Division of Mineralogy, University of\\
%Salzburg, Hellbrunnerstr. 34, Salzburg A-5020, Austria}
\date{\today}

\begin{abstract}
We report the results of a $^{45}$Sc nuclear magnetic resonance (NMR) study
on the quasi-one-dimensional compound Cu$_2$Sc$_2$Ge$_4$O$_{13}$ at
temperatures between 4 and 300 K. This material has been a subject of
current interest due to indications of spin gap behavior. The
temperature-dependent NMR shift exhibits a character of low-dimensional
magnetism with a negative broad maximum at $T_{max}$ $\simeq $ 170 K. Below $%
T_{max}$, the NMR shifts and spin lattice relaxation rates clearly indicate
activated responses, confirming the existence of a spin gap in Cu$_2$Sc$_2$Ge%
$_4$O$_{13}$. The experimental NMR data can be well fitted to the spin dimer
model, yielding a spin gap value of about 275 K which is close to the 25 meV
peak found in the inelastic neutron measurement. A detailed analysis further
points out that the nearly isolated dimer picture is proper for the
understanding of spin gap nature in Cu$_2$Sc$_2$Ge$_4$O$_{13}$.
\end{abstract}

\pacs{76.60.-k, 75.10.Pq}

\maketitle

\section{Introduction}

The physics of low-dimensional magnetic systems continues to attract
attention because of the association with peculiar quantum effects.\cite
{Lemmens:03} Prominent examples like the spin-Peierls transition and the
spin ladder compounds have been characterized by ground states of the spin
singlet with a finite spin gap.\cite{CuGeO:93,16} For a
quasi-one-dimensional chain with half-integer spin ($S$ = 1/2), the gap can
be opened via either frustration due to next nearest neighbor
antiferromagnetic exchange or dimerization due to an alternating coupling to
nearest neighbors along the chain.\cite{2,3} Several $S$ = 1/2 chain systems
such as (VO)$_2$P$_2$O$_7$, BaCu$_2$V$_2$O$_8$, and PbCu$_2$(PO$_4$)$_2$
have been reported to possess spin gaps and their characteristics have been
interpreted in accordance with these scenarios.\cite{V:94,He:04,A:06}

Cu$_2$Sc$_2$Ge$_4$O$_{13}$, which crystallizes in a monoclinic structure
with the space group {\it P}2$_1${\it /m}, was recently synthesized by one
of the present authors (G.J.R.).\cite{R:04} A schematic picture of the
crystal structure is illustrated in Fig. 1. Taking into account the known
oxidation states of O$^{2-}$ and Ge$^{4+}$, the remaining valences are
nonmagnetic Sc$^{3+}$ and Cu$^{2+}$ ($S$ = 1/2). The spin chain in Cu$_2$Sc$%
_2$Ge$_4$O$_{13}$ can be described as an arrangement of spin dimers oriented
along the crystallographic {\it b}-axis, separated by GeO$_4$ tetrahedra and
crankshaft-like chains of ScO$_6$ octahedra (Fig. 1). Each dimer consists of
two Cu$^{2+}$ ions in a Cu$_2$O$_2$ plaquette and weak interdimer
interactions could be possible to induce a quantum phase transition from a
gapless state into a gapped state for Cu$_2$Sc$_2$Ge$_4$O$_{13}$.

The bulk magnetic susceptibility of Cu$_2$Sc$_2$Ge$_4$O$_{13}$ exhibits a
broad maximum at around 170 K and decreases rapidly at low temperatures,
indicative of spin gap behavior for this material.\cite{Ma:06} Also the
magnetization data have been fitted well to the dimer chain model, yielding
a spin gap of 290 K (25 meV){\it .} Furthermore, the result of heat capacity
has confirmed no magnetic ordering above 2 K.\cite{Ma:06} Recently, Masuda
{\it et al.} have performed a neutron inelastic-scattering experiment on Cu$%
_2$Sc$_2$Ge$_4$O$_{13}$ and found a narrow-band excitation of about 25 meV,
which is quite consistent with the value extracted from the susceptibility.%
\cite{Ma:06} They thus attributed the observation to the spin-gap excitation
and proposed Cu$_2$Sc$_2$Ge$_4$O$_{13}$ as a new spin-gapped dimer material.
%%%%%%%%%%%%%%%%%%%%%%%%%%%%%%%%%%%%%%%%%%%%%%%%%%%%%%%%%%%%%%%%%%%%%%%%%
\begin{figure}[htbp]
\includegraphics*[width=3 in]{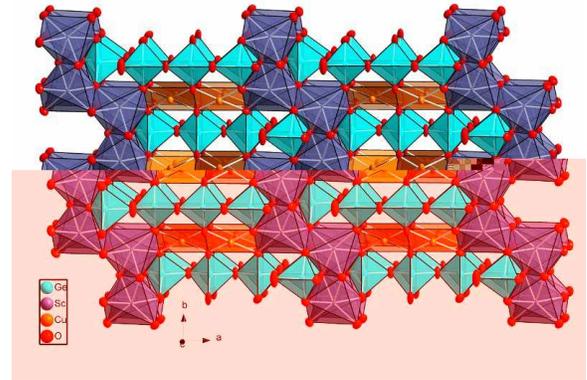}
\caption{Crystal structure for Cu$_2$Sc$_2$Ge$_4$O$_{13}$.}
 \label{fig:fig1c}
\end{figure}
%%%%%%%%%%%%%%%%%%%%%%%%%%%%%%%%%%%%%%%%%%%%%%%%%%%%%%%%%%%%%%%%%%%%%%%%

In order to further identify the existence of a spin gap in Cu$_2$Sc$_2$Ge$%
_4 $O$_{13}$, we carried out a detailed $^{45}$Sc nuclear magnetic resonance
(NMR) study invoking NMR shifts and spin-lattice relaxation rates on this
compound. The NMR shift provides a local measurement of the susceptibility
which is less sensitive to the presence of impurities and other phases. The
spin-lattice relaxation rate is a sensitive probe for the low-energy spin
excitations, yielding direct evidence for the presence of an energy gap.
Regarding the title compound Cu$_2$Sc$_2$Ge$_4$O$_{13}$, a transfer of
magnetic {\it d}-spin from the Cu$^{2+}$ onto Sc$^{3+}$ 4{\it s} orbital
allows us to probe the Cu$^{2+}$ spin dynamics and determine the spin gap
through the transfer hyperfine interaction. The experimental NMR results
clearly reveal spin gap behavior in this material. Data analysis using the
isolated dimer model is found to give good agreement with the observations.
Interestingly, the deduced spin gap value of 275 K is close to the 25 meV
peak revealed by the inelastic neutron experiment,\cite{Ma:06} indicating
the same energy excitation detected by both measurements.

\section{Experiments and Discussion}

A polycrystalline Cu$_2$Sc$_2$Ge$_4$O$_{13}$ sample was synthesized by a
ceramic sintering solid-state reaction technique.\cite{R:04} A mixture of
CuO, Sc$_2$O$_3$ and GeO$_2$, weighted in a predetermined molar ration of Cu$%
_2$Sc$_2$Ge$_4$O$_{13}$, was carefully ground under alcohol, pressed into
pellets, put into an open platinum crucible, and fired under ambient
pressure and ambient oxygen fugacity in a temperature range between 900 $%
^{\circ }$C and 1100 $^{\circ }$C. After each of six heating cycle, the
sample was reground, pressed, and reheated. In initial stages the product
consisted of a mixture of the title compound and CuGeO$_3$. The amount of
CuGeO$_3$ reduced successively as increasing synthesis temperature and
heating time. In a final synthesis cycle, the sample was fired at 1150 $%
^{\circ }$C and a light blue product was thus obtained. A room-temperature
x-ray diffraction confirmed the single phase for Cu$_2$Sc$_2$Ge$_4$O$_{13}$
with the lattice parameters $a=$12.336(2) $\stackrel{\circ }{A}$, $b=$%
8.7034(9) $\stackrel{\circ }{A}$, $c=$4.8883(8) $\stackrel{\circ }{A}$, and $%
\beta =$95.74(2).

NMR experiments were performed using a Varian 300 spectrometer, with
a constant field of 7.05 T. A home-built probe was employed for the
low-temperature measurements.\cite{Lue:06} The powder specimen was
put in a plastic vial that showed no observable $^{45}$Sc NMR
signal. The NMR spectrum was obtained from spin echo fast Fourier
transforms using a standard $\pi /2-\tau -\pi $ sequence. Within the
{\it P}2$_1${\it /m} space group for Cu$_2$Sc$_2$Ge$_4$O$_{13}$,
scandium atoms occupy one crystallographic site having octahedral
oxygen coordination, yielding an one-site $^{45}$Sc NMR powder
pattern, as demonstrated in the inset of Fig. 2. Here the observed
NMR line shape is manifested by the anisotropic Knight shift as well
as the quadrupole effects. However, no visible quadrupole edges have
been detected with our static NMR probe, suggesting that the value
of quadrupole frequency $\nu _Q$ is essentially small. Such a result
indicates that the Sc atoms are located in a symmetric environment,
probably at the center of the ScO$_6$ octahedron. It is of great
importance to note that the line shape of the spectrum remains
unchanged with temperature. This
phenomenon confirms the nonmagnetic ground state for Cu$_2$Sc$_2$Ge$_4$O$%
_{13}$ and no structural changes above 4 K.
%%%%%%%%%%%%%%%%%%%%%%%%%%%%%%%%%%%%%%%%%%%%%%%%%%%%%%%%%%%%%%%%%%%%%%%%%
\begin{figure}[htbp]
\includegraphics*[width=2.5 in]{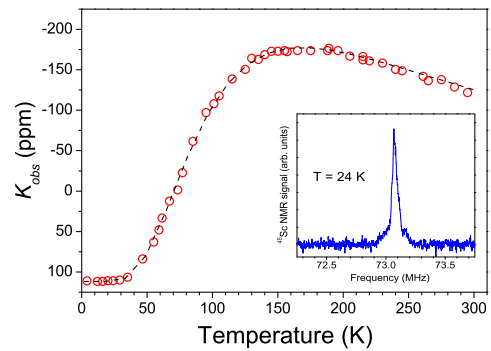}
\caption{Temperature dependence of the observed $^{45}$Sc NMR shift
in Cu$_2$Sc$_2$Ge$_4$O$_{13}$. Note that $K_{obs}$ is shown by the
negative direction. Dashed curve: fit to the dimer model plus a
constant term. Inset: $^{45}$Sc NMR spectrum for
Cu$_2$Sc$_2$Ge$_4$O$_{13}$ measured at 24 K.}
 \label{fig:fig2c}
\end{figure}
%%%%%%%%%%%%%%%%%%%%%%%%%%%%%%%%%%%%%%%%%%%%%%%%%%%%%%%%%%%%%%%%%%%%%%%%

In Fig. 2, we displayed the observed temperature-dependent NMR shift ({\it K}%
$_{obs}$) for the Sc site. The shift was taken at the peak position of the
resonance line referred to the $^{45}$Sc resonance frequency of aqueous ScCl$%
_3$. Here {\it K}$_{obs}$ has different sign from that of the susceptibility
since the transfer hyperfine field arising from 3{\it d} electrons of Cu$%
^{2+}$ ions is negative. In spite of the sign difference of {\it K}$_{obs}$,
the whole temperature variation is quite consistent with the bulk
susceptibility data, with a broad maximum at around $T_{max}$ $\simeq $ 170
K and a rapid decrease as lowering temperature.\cite{Ma:06} In general, {\it %
K}$_{obs}$ can be decomposed into {\it K}$_{obs}$ = {\it K}$_o+$ {\it K}$%
_{spin}$({\it T}). The first term {\it K}$_o$ = 112 ppm, mainly arising from
the orbital shift, is independence of temperature. On the other hand, the
spin shift {\it K}$_{spin}$, which reflects the Cu$^{2+}$ spin behavior
through the transfer hyperfine interaction, is a function of temperature.
%%%%%%%%%%%%%%%%%%%%%%%%%%%%%%%%%%%%%%%%%%%%%%%%%%%%%%%%%%%%%%%%%%%%%%%%%
\begin{figure}[htbp]
\includegraphics*[width=2.5 in]{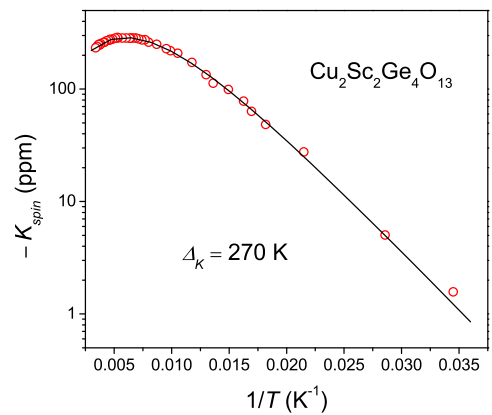}
\caption{Plot of $K_{spin}$ against 1/{\it T} for Cu$_2$Sc$_2$Ge$_4$O$%
_{13}$. Solid curve: fit to the function described by the dimer model with $%
\Delta _K$ of 270 K.}
 \label{fig:fig3c}
\end{figure}
%%%%%%%%%%%%%%%%%%%%%%%%%%%%%%%%%%%%%%%%%%%%%%%%%%%%%%%%%%%%%%%%%%%%%%%%

As mentioned above, the spin chain configuration of spin dimers
could be responsible for the magnetic behavior of
Cu$_2$Sc$_2$Ge$_4$O$_{13}$. Accordingly, the temperature dependence
of the spin shift will obey the relation {\it K}$_{spin}$ $\propto $
1/$T$($3+e^{\Delta _K/T}$).\cite {Troyer:94} Here $\Delta _K$ is the
gap energy determined from the NMR shift based on the dimer model.
As demonstrated in Fig. 3, {\it K}$_{spin}$ can be fitted well in a
fairly wide temperature range to this relation (solid curve in Fig.
3). It is also found that {\it K}$_{spin}$ vanishes to almost zero
at low temperatures, indicative of the gap in the spin excitation
spectrum. With this fit, we extracted $\Delta _K$ = 270$\pm $20 K,
consistent with 290 K obtained from the susceptibility
measurement.\cite{Ma:06}

The spin shift here is related to the magnetic susceptibility $\chi
$ by the expression
\begin{equation}
K_{spin}(T)=\frac{H_{hf}^{tr}}{N_A\mu _B}\chi (T),
\end{equation}
where {\it H}$_{hf}^{tr}$ is the transfer hyperfine field due to an
intermixing of Sc and Cu spin states, $N_A$ is the Avogadro's constant, and $%
\mu _B$ is the Bohr magneton. The Clogston-Jaccarino plot\cite{Cl:64} which
shows the observed shift against magnetic susceptibility (deduced from the
data of Ref. 10) is given in Fig. 4. The linear behavior indicates a unique
hyperfine field over the entire temperature range we investigated. The slope
yields a small value of {\it H}$_{hf}^{tr}$ = $-$0.66$\pm $0.04 kOe.
%%%%%%%%%%%%%%%%%%%%%%%%%%%%%%%%%%%%%%%%%%%%%%%%%%%%%%%%%%%%%%%%%%%%%%%%%
\begin{figure}[htbp]
\includegraphics*[width=2.1 in]{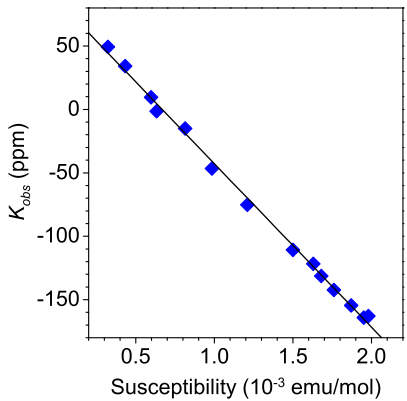}
\caption{Variation of $K_{obs}$ versus $\chi $ (deduced from Ref.
10)}
 \label{fig:fig4c}
\end{figure}
%%%%%%%%%%%%%%%%%%%%%%%%%%%%%%%%%%%%%%%%%%%%%%%%%%%%%%%%%%%%%%%%%%%%%%%%

To gain more insight into the spin gap characteristics of Cu$_2$Sc$_2$Ge$_4$O%
$_{13}$, we performed spin-lattice relaxation rate (1/{\it T}$_1$)
measurements, being sensitive to the low-energy magnetic excitations. It
thus provides direct information about the low-energy spin dynamics and the
presence of a spin gap. Here the {\it T$_1$} measurement was carried out
using the saturation recovery method. The saturation rf comb with 30 short 2
$\mu $s pulses was employed. We recorded the recovery of the signal strength
by integrating the $^{45}$Sc spin echo signal. For the central transition
with {\it I} = 7/2, the recovery of the nuclear magnetization follows\cite
{Simmons:62}

\begin{equation}
\frac{M(t)}{M(0)}=0.012e^{-\frac t{T_1}}+0.068e^{-\frac{6t}{T_1}}+0.206e^{-%
\frac{15t}{T_1}}+0.714e^{-\frac{28t}{T_1}}.
\end{equation}
Here $M$($t$) is the magnetization at the recovery time $t$ and $M$($0$) is
the initial magnetization. The {\it T$_1$} value was thus obtained by
fitting to this multi-exponential function. To provide accurate values, each
{\it T$_1$} has been measured several times and the averaged {\it T$_1$} for
the corresponding temperature is shown in the inset of Fig. 5. It is
apparent that 1/{\it T$_1$ }exhibits activated behavior at low temperatures.
Based on the dimerized scenario, the spin-lattice relaxation rate should be
fitted to the form 1/{\it T}$_1$ $\propto $1/$(3+e^{\Delta _R/T})$ by
analogy to the treatment of NMR shift.\cite{Troyer:94} The fitting result,
drawn in Fig. 5 as a solid curve, is quite satisfactory and yields an energy
gap $\Delta _R$ = 275$\pm $25 K. Remarkably, this excitation energy is
almost identical with the 25 meV (290 K) peak obtained in the inelastic
neutron experiment,\cite{Ma:06} pointing to the same energy excitation
probed by both measurements.
%%%%%%%%%%%%%%%%%%%%%%%%%%%%%%%%%%%%%%%%%%%%%%%%%%%%%%%%%%%%%%%%%%%%%%%%%
\begin{figure}[htbp]
\includegraphics*[width=2.2 in]{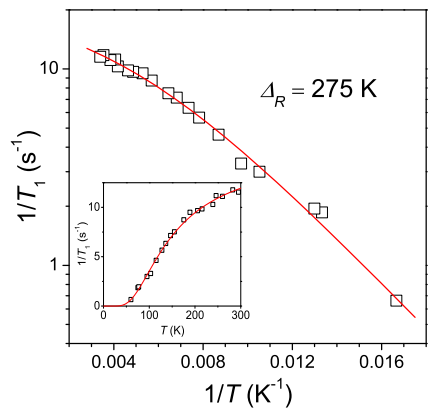}
\caption{Inverted temperature dependence of 1/$T_1$ for
Cu$_2$Sc$_2$Ge$_4$O$_{13}$. The solid curve is the fitted function
based on the dimer model. Inset shows the temperature variation of
1/$T_1$.}
 \label{fig:fig5c}
\end{figure}
%%%%%%%%%%%%%%%%%%%%%%%%%%%%%%%%%%%%%%%%%%%%%%%%%%%%%%%%%%%%%%%%%%%%%%%%

Our NMR investigation thus provides clear evidence for the existence
of spin gap in Cu$_2$Sc$_2$Ge$_4$O$_{13}$. The extracted $\Delta _K$
was found to be close to $\Delta _R$. As a matter of fact, the ratio
of $\Delta _R$/$\Delta _K$ $\simeq $ 1 is commonly seen in the
dimerized systems.\cite {Furukawa:96,15} Within the isolated dimer
limit, $T_{max}$, the temperature at which magnetic susceptibility
or NMR shift display a maximum, will appear
at around 0.63$\Delta _R$.\cite{Itoh:97} For the present case of Cu$_2$Sc$_2$%
Ge$_4$O$_{13}$, the determined $\Delta _R$ = 275 K results in $T_{max}$ =
173 K, in excellent agreement with the observed $T_{max}$ $\simeq $ 170 K.
In this regard, it seems reasonable to characterize Cu$_2$Sc$_2$Ge$_4$O$_{13}
$ as a nearly isolated dimer chain compound.
%%%%%%%%%%%%%%%%%%%%%%%%%%%%%%%%%%%%%%%%%%%%%%%%%%%%%%%%%%%%%%%%%%%%%%%%%
\begin{figure}[htbp]
\includegraphics*[width=2.5 in]{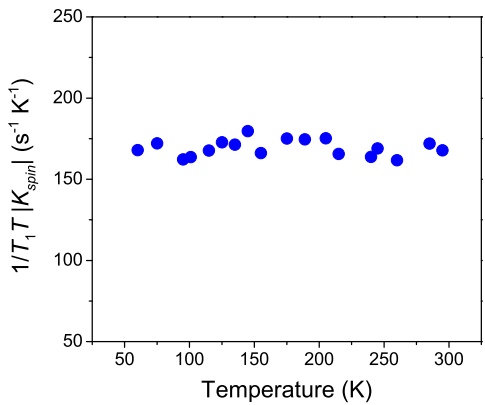}
\caption{Temperature variation of ($T_1T$)$^{-1}$/$\mid K_{spin}\mid
$.}
 \label{fig:fig6c}
\end{figure}
%%%%%%%%%%%%%%%%%%%%%%%%%%%%%%%%%%%%%%%%%%%%%%%%%%%%%%%%%%%%%%%%%%%%%%%%

At this moment, we did not attempt to fit the temperature dependence
of spin shift and 1/{\it T$_1$} to other models such as the
alternating chain model. In fact, our previous analyses of NMR\ data
for BaCu$_2$V$_2$O$_8$ indicated that both alternating chain and
dimer chain models are suitable for the understanding of the spin
gap nature.\cite{Lue:05} However, a convincing result can be
established unambiguously from the comparison of static and dynamic
excitations. Within the dimer picture, the excitation is predominant
by a simple singlet-triplet process.\cite{D:88} As a result, the
static susceptibility probed by $K_{spin}$ and the local dissipative
susceptibility probed by ($T_1T$)$^{-1}$ are nearly identical. On
the other hand, for weakly coupled Heisenberg spin chains, both
triplet and singlet-triplet mechanisms contribute to the relaxation
rate.\cite{She:96} In this case, the temperature dependence of
($T_1T$)$^{-1}$ becomes more prominent than that
of $K_{spin}$ as $T<$ $\Delta _K$, leading to a larger spin gap deduced by 1/%
{\it T$_1$.}\cite{18,Kishine:97,19,20} Such a phenomenon has been observed
in the $S=$1 Haldane gap compound Y$_2$BaNiO$_5$ and the $S=$1/2 spin chain
BaCu$_2$V$_2$O$_8$.\cite{Sh:95,Ch:05} To provide a reliable spin gap
description for Cu$_2$Sc$_2$Ge$_4$O$_{13}$, we thus examine the temperature
dependence of ($T_1T$)$^{-1}$/$\mid K_{spin}\mid $ without any fitting. As
one can see from Fig. 6, the ratio of ($T_1T$)$^{-1}$/$\mid K_{spin}\mid $
remains constant even below the spin gap. This result is consistent with the
spin dimer theory in which both static and dynamic excitations follow the
same temperature variation due to the involvement of a simple
singlet-triplet excitation.

\section{Conclusions}

In conclusion, we report the first NMR investigation of Cu$_2$Sc$_2$Ge$_4$O$%
_{13}$ and present evidence for the existence of spin gap in this
material. The spin gap of about 275 K deduced from the spin shifts
and spin-lattice relaxation rates was found to be identical. A
detailed analysis further indicates that the spin gap nature can be
well accounted for by the free dimer model. These findings thus
allow us to add Cu$_2$Sc$_2$Ge$_4$O$_{13}$ to the family of $S=$1/2
dimerized spin gap compounds.\\

\section*{ ACKNOWLEDGEMENTS}

This work was supported by the National Science Council of Taiwan under
Grant No. NSC-95-2112-M-006-021-MY3 (C.S.L), and by a grant to G.J.R. from
the {\it Fond zur Forderung der Wissenschaftlichen Forschung }(FWF),
Austria, (Grant No. R33-N10).

% now the references. delete or change fake bibitem. delete next three
%   lines and directly read in your .bbl file if you use bibtex.

\end{document}